\documentclass[english,10pt,aps,pra,twocolumn,floatfix]{revtex4-1}

\usepackage{amsmath}
\usepackage{amssymb}
\usepackage{amsfonts}
\usepackage{bm}
\usepackage{bbm}
\usepackage{epsfig}
\usepackage{grffile}
\usepackage{graphics}

\usepackage[usenames,dvipsnames]{color}
\definecolor{dblue}{rgb}{0,0,.5}

\usepackage[colorlinks=true,citecolor=dblue,linkcolor=dblue,urlcolor=dblue]{hyperref}
\usepackage[all]{hypcap}
\newcommand{\Footnote}[1]{\footnote{\unexpanded{#1}}}

\newcommand{\ud}{\mathrm{d}}
\newcommand{\id}{\mathbbm{1}}
\newcommand{\hid}{\hat{\mathbbm{1}}}
\newcommand{\Tr}{\operatorname{Tr}}
\newcommand{\bra}{\langle}
\newcommand{\ket}{\rangle}
\newcommand{\mc}[1]{\mathcal{#1}}

\renewcommand{\H}{\mc{H}}
\newcommand{\Hb}{\bar{\mc{H}}}
\newcommand{\hH}{\hat{H}}
\newcommand{\hS}{\hat{S}}
\newcommand{\hU}{\hat{U}}
\newcommand{\hX}{\hat{X}}
\newcommand{\hY}{\hat{Y}}
\newcommand{\hvS}{\hat{\vec{S}}}
\newcommand{\dm}{{\hat{\varrho}}}
\newcommand{\dmt}{{\tilde{\varrho}}}
\newcommand{\dmp}{\varrho}
\newcommand{\Dm}{{\hat{\Omega}}}
\renewcommand{\vec}[1]{{\boldsymbol{#1}}}

\newcommand{\vs}{{\vec{\sigma}}}
\newcommand{\vsb}{{\bar{\vec{\sigma}}}}
\newcommand{\sigmab}{\bar{\sigma}}
\newcommand{\vphi}{\varphi}
\newcommand{\omegat}{\tilde{\omega}}
\newcommand{\Tw}{\mathcal{T}}
\newcommand{\CC}{\mathbb{C}}
\newcommand{\E}{\mc{E}}
\newcommand{\aux}{\text{aux}}
\newcommand{\SVD}{\text{\tiny SVD}}

\newcommand{\duke} {Department of Physics, Duke University, Durham, North Carolina 27708, USA}

\newcommand{\Title} {One-dimensional quantum systems at finite temperatures can be simulated efficiently on classical computers}
\newcommand{\Authors}
{
\author{Thomas Barthel}
\affiliation{\duke}
}
\newcommand{\Date} {July 27, 2017}

\begin{document}

\title{\Title\texorpdfstring{\vspace{0.5em}}{}}
\Authors

\begin{abstract}
It is by now well-known that ground states of gapped one-dimensional (1d) quantum-many body systems with short-range interactions can be studied efficiently using classical computers and matrix product state techniques. A corresponding result for finite temperatures was missing.
Using the replica trick in 1+1d quantum field theory, it is shown here that the cost for the classical simulation of 1d systems at finite temperatures grows in fact only polynomially with the inverse temperature and is system-size independent -- even for gapless systems. In particular, we show that the thermofield double state (TDS), a purification of the equilibrium density operator, can be obtained efficiently in matrix-product form. The argument is based on the scaling behavior of R\'{e}nyi entanglement entropies in the TDS. At finite temperatures, they obey the area law. For gapless systems with central charge $c$, the entanglement is found to grow only logarithmically with inverse temperature, $S_\alpha\sim \frac{c}{6}(1+1/\alpha)\log \beta$.
The field-theoretical results are tested and confirmed by quasi-exact numerical computations for integrable and non-integrable spin systems, and interacting bosons.\vspace{3em}
\end{abstract}

\date{\Date}
\maketitle

\section{Introduction}\label{sec:intro}
The number of degrees of freedom of a quantum many-body system grows exponentially with its size. This makes it often difficult to study these systems analytically or numerically, but has also lead to the promising concepts of quantum computation and simulation. A focus of current research is to identify realms of the quantum world that can or cannot be captured efficiently using classical computers. Entanglement properties of the systems allow us to quantify their complexity. It has been established \cite{Holzhey1994-424,Vidal2003-7,Calabrese2004,Verstraete2005-5,Hastings2007-08} that ground states of typical one-dimensional (1d) systems can be approximated by matrix product states (MPS) \cite{Accardi1981,Affleck1987,Fannes1992-144,Rommer1997}, which are efficient classical representations. This is brought to bear in the density-matrix renormalization group method (DMRG) \cite{White1992-11,White1993-10,Schollwoeck2005}. In a more recent breakthrough, it has been proven that such MPS approximations of ground states can indeed be computed efficiently and certifiably for gapped local 1d systems \cite{Landau2015-11}.

For typical 1d systems at finite temperatures, we investigate here the scaling behavior of R\'{e}nyi entanglement entropies $S_\alpha$ in the thermofield double state (TDS) \cite{Takahashi1975-2,Israel1976-57}
\begin{equation}\label{eq:TDS}
	|\dmp_\beta\ket=\frac{e^{-\beta\hH/2}\otimes\hid}{\sqrt{Z}}|\id\ket\quad\text{with} \ \
	|\id\ket=\sum_{\vs}|\vs\ket\otimes|\vs\ket
\end{equation}
which is a purification \cite{Uhlmann1976,Uhlmann1986,Nielsen2000} of the equilibrium density operator $\dm_\beta=\exp(-\beta\hH)/Z$, where $\beta={1}/{k_BT}$ is the inverse temperature, $\mc{B}=\{|\vs\ket\}$ is an orthonormal basis for the many-body Hilbert space $\H$, and $Z=\Tr[\exp(-\beta\hH)]$ is the partition function. The results imply that an efficient classical simulation is possible \emph{at any finite temperature}.
\begin{figure*}[t]
\includegraphics[width=\textwidth]{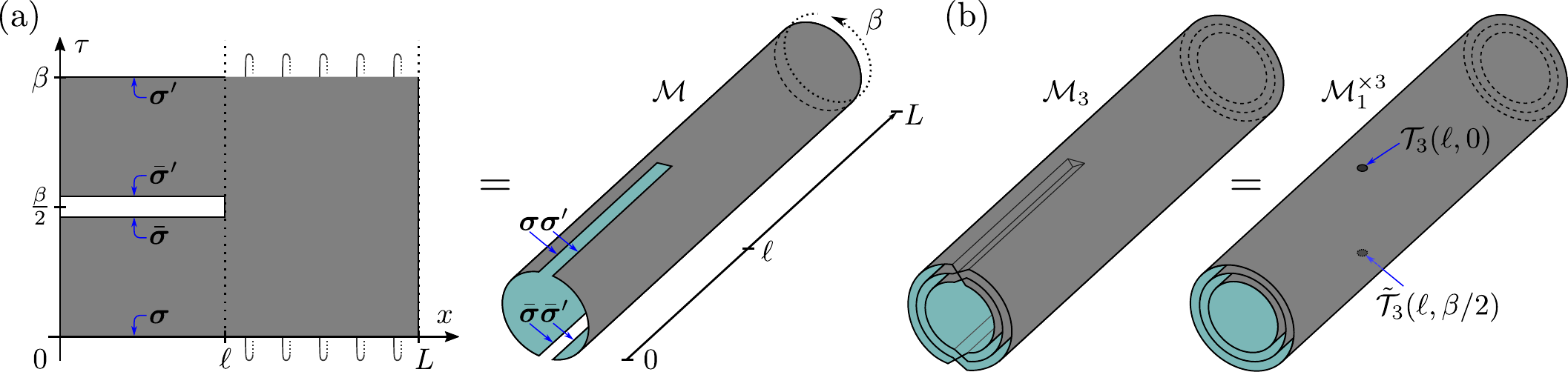}
\caption{\label{fig:Cylinders}\textbf{Field-theoretical evaluation of TDS entanglement.} In the evaluation of $S_\alpha$ [Eq.~\eqref{eq:Renyi}] for the TDS $|\dmp_\beta\ket$, imaginary time $\tau$ is introduced through Trotter decomposition of $\exp(-\beta\hH)$.
(a) The reduced density matrix $\bra\vs\vsb|\Dm_A|\vs'\vsb'\ket$ for subsystem $A$ with $x\leq \ell$ corresponds to an Euclidean path integral on a cylinder $\mc{M}$ of length $L$ and circumference $\beta$ with states $\vs$, $\vs'$, $\vsb$, and $\vsb'$ at two parallel cuts of length $\ell$ at $\tau=0$ and $\tau=\beta/2$.
(b) According to the replica trick for integer $\alpha=n$, the Euclidean path integral \eqref{eq:Zn} for $Z_n$ is evaluated on $n$ cyclically connected copies $\mc{M}_n$ of the cylinder with branch points at $(x,\tau)=(\ell,0),(\ell,\beta/2)$. It turns out to be a simple two-point correlator \eqref{eq:Tw} of branch-point twist fields $\Tw_n$ and $\tilde{\Tw}_n$ on $n$ unconnected cylinders $\mc{M}_1^{\times n}$.}
\end{figure*}

We determine the scaling of entanglement entropies for critical systems described by 1+1d conformal field theory (CFT) and for gapped systems described by massive relativistic quantum field theory. This, together with arguments from Ref.~\cite{Verstraete2005-5}, leads to the conclusion that thermal equilibrium states of these systems can be represented faithfully as MPS in the sense that $|\dmp_\beta\ket\approx |\dmp_\beta^D\ket$, where
\begin{equation}\label{eq:TDS-MPS}
	|\dmp_\beta^D\ket:=\sum_{\vs,\vsb\in\mc{B}}A^{\sigma_1,\sigmab_1}_1 A^{\sigma_2,\sigmab_2}_2\dotsb A_L^{\sigma_L,\sigmab_L}\,|\vs\ket\otimes|\vsb\ket
\end{equation}
with $|\vs\ket\equiv|\sigma_1,\dotsc,\sigma_L\ket$ for a system with $L$ lattice sites, $D\times D$ matrices $A^{\sigma_i,\sigmab_i}_i$, and corresponding row and column vectors $A^{\sigma_1,\sigmab_1}_1$ and $A^{\sigma_L,\sigmab_L}_L$. For a fixed approximation accuracy, the so-called bond dimension $D$ does not have to be increased with increasing system size $L$. $D$ is also temperature-independent for gapped systems, $D=\mc{O}(L^0\beta^0)$, and scales only polynomially in the inverse temperature for gapless systems, $D=\mc{O}(L^0\beta^{\lambda})$ with exponent $\lambda=\frac{c}{6}(1+1/\alpha)$ as discussed below. This allows for an efficient simulation on classical computers, using time-dependent DMRG methods \cite{Vidal2003-10,White2004,Daley2004,Orus2008-78}. Our results are in this sense a finite-temperature counterpart of Refs.~\cite{Holzhey1994-424,Vidal2003-7,Calabrese2004,Verstraete2005-5,Hastings2007-08,Landau2015-11}.
As the obtained bounds on $D$ are relatively loose, we support the analytical results with quasi-exact numerical simulations for integrable and non-integrable systems.

Employing a truncated cluster expansion of $\exp(-\beta\hH)$ for lattice models with norm-bounded short-range interactions, Hastings proved earlier \cite{Hastings2006-73} that matrix product approximations are possible, albeit with a considerably worse scaling behavior for $D$. In his construction, above a certain percolation temperature, the required bond dimension grows polynomially in the system size, $D=e^{\mc{O}(\log L)}$, and exponentially in $\beta$ for temperatures below the percolation temperature, $D=e^{\mc{O}(\beta\log L)}$ \cite{Hastings2006-73,Kliesch2014-04,Molnar2015-91}. Our field-theoretical and numerical findings, hence, substantially improve over these results.

\section{Purification of mixed states}\label{sec:Purify}
For every density operator $\dm$ on a Hilbert space $\H$, there exist pure states $|\dmp\ket$ on an enlarged Hilbert space $\H\otimes\Hb$ that \emph{purify} $\dm$ in the sense that $\Tr_{\Hb}|\dmp\ket\bra\dmp|=\dm$ \cite{Uhlmann1976,Uhlmann1986,Nielsen2000}, where the partial trace is over the second component of the tensor product space \Footnote{$\Tr_{\Hb}\hX:=\sum_{\vs,\vs',\vsb}|\vs\ket\bra\vs\vsb|\hX|\vs'\vsb\ket\bra\vs'|$ for $\hX$ on $\H\otimes\Hb$.}. Hence, $|\dmp\ket$ contains the full information about $\dm$ and expectation values can be evaluated according to
$\Tr(\dm\hat{O})= \bra\dmp|\hat{O}\otimes \hid|\dmp\ket$.
One possible choice for such a purification, is to use the \emph{vectorization} of $\dm^{1/2}$, i.e., $|\dmp\ket:=\sum_{\vs,\vsb\in\mc{B}}\bra\vs|\dm^{1/2}|\vsb\ket\cdot|\vs\ket\otimes|\vsb\ket\in\H\otimes\Hb$ with $\Hb=\H$. For thermal states $\dm_\beta$, this choice leads to the TDS \eqref{eq:TDS}.
This specific purification of $\dm_\beta$ is extensively used in MPS simulations of (strongly-correlated) systems at finite temperatures. See, for example, Refs.~\cite{Feiguin2005-72,Barthel2005,Ruegg2008-101,Barthel2009-79b,Feiguin2010-81,Bouillot2011-83,Karrasch2012-108,Barthel2012_12,Barthel2013-15,Karrasch2013-87,Lake2013-111,Tiegel2014-90,Karrasch2014-90,Gori2016-93,Tiegel2016-93,Coira2016-94}. Starting at infinite temperature with $|\id\ket$, $|\dmp_\beta\ket$ is obtained through an imaginary-time evolution \cite{Verstraete2004-6,Schollwoeck2011-326,Barthel2016-94}.

\section{Entanglement entropies}\label{sec:EE}
The scaling behavior of R\'{e}nyi entanglement entropies $S_\alpha$ in the purifications $|\dmp_\beta\ket\in\H\otimes\Hb$ will allow us to bound bond dimensions $D$ needed to approximate $|\dmp_\beta\ket$ to a certain precision in MPS form \eqref{eq:TDS-MPS}. Specifically, we consider a (spatial) bipartition of the system such that $\H=\H_A\otimes\H_B$ and, correspondingly, $\Hb=\Hb_A\otimes\Hb_B$. Then,
\begin{equation}\label{eq:Renyi}
	S_\alpha(A):=\frac{1}{1-\alpha}\log\Tr\Dm_A^\alpha,
	\quad
	\Dm_A:=\Tr_{\H_B,\Hb_B}|\dmp_\beta\ket\bra\dmp_\beta|.
\end{equation}
$S_\alpha$ is a generalization of the famous von Neumann entanglement entropy which is recovered for $\alpha\to 1$.
Note that $S_\alpha(A)$ has no immediate relation to the entanglement of purification \cite{Terhal2002}.

\section{Entanglement from 1+1d field theory}\label{sec:QFT}
Choosing subsystem $A$ to consist of sites $[1,\ell]$ and introducing imaginary time $\tau$ through a Trotter decomposition $e^{-\beta\hH}=(e^{-\frac{\beta}{M}\hH})^M$, the reduced density matrices $\bra\vs\vsb|\Dm_A|\vs'\vsb'\ket$ of the TDS correspond to an Euclidean path integral on a cylinder $\mc{M}$ of length $L$ and circumference $\beta$ with two parallel cuts of length $\ell$ at distance $\Delta\tau=\beta/2$ \Footnote{We use natural units where Hamiltonian $\hH$ and $\beta^{-1}$ have been divided by a characteristic energy scale such that, at criticality, gapless excitations propagate at speed 1.}. The four open lines represent the basis states $|\vs\vsb\ket,\,|\vs'\vsb'\ket\in\H_A\otimes\Hb_A$ as indicated in Fig.~\ref{fig:Cylinders}a. To evaluate the R\'{e}nyi entanglement entropies \eqref{eq:Renyi} for integer $\alpha=n\in\mathbb{N}$, we can employ the \emph{replica trick} \cite{Holzhey1994-424,Callan1994-333} and compute the required ratio of partition functions
	${Z_n}/{Z^n}:=\Tr\Dm_A^n=\sum_{\{\vs_k\},\{\vsb_k\}}\bra\vs_1\vsb_1|\Dm_A|\vs_2\vsb_2\ket\dots\bra\vs_n\vsb_n|\Dm_A|\vs_1\vsb_1\ket$
by evaluating the Euclidean path integral on the manifold $\mc{M}_n$, consisting of $n$ cylinders which are cyclically connected along the described cuts. Then,
\begin{equation}\label{eq:Zn}
	Z_n=\int\mc{D}[\vphi_1,\dotsc,\vphi_n]_{\mc{M}_n} e^{-\int_{\mc{M}}\ud x\ud \tau \sum_{k=1}^n\mc{L}[\vphi_k](x,\tau)},
\end{equation}
where we have made the transition to a 1+1d continuum field theory with Lagrange density $\mc{L}$, describing the long-range properties of the system, and $\vphi_k(x,\tau)$ is the field on the $k$\textsuperscript{th} replica of the cylinder $\mc{M}$. Continuity conditions appropriate for $\mc{M}_n$ are imposed. For $x>\ell$, $\mc{M}_n$ consists of $n$ unconnected cylinders of circumference $\beta$, just as in the computation of ground-state R\'{e}nyi entanglement entropies \cite{Calabrese2004}. But a decisive difference is that, in our case, $\mc{M}_n$ also consists of $n$ unconnected cylinders for $x\leq\ell$, each being composed of one half of replica $k$ and one half of replica $k+1\mod n$ with $1\leq k\leq n$. As we will see explicitly, this is the reason why entanglement entropies of finite-temperature TDS \eqref{eq:TDS} obey the area law (saturate with increasing $\ell$), while they are $\propto\log\ell$ for ground states of critical systems \cite{Holzhey1994-424,Callan1994-333,Vidal2003-7,Jin2004-116,Calabrese2004,Zhou2005-12}.

Following Refs.~\cite{Knizhnik1987-112,Cardy2008-130}, the partition function \eqref{eq:Zn} for the non-trivial manifold $\mc{M}_n$ can be expressed as a correlation function of \emph{branch-point twist fields} $\Tw_n$ and $\tilde{\Tw}_n$ such that, in our case,
\begin{equation}\label{eq:TwDef}
	Z_n/Z^n=\big\bra\tilde{\Tw}_n(\ell,{\beta}/{2})\,\Tw_n(\ell,0)\big\ket_{\mc{M}_1^{\times n}}
\end{equation}
with $\bra\dots\ket_{\mc{M}_1^{\times n}}$ denoting the expectation value according to the path integral over $n$ unconnected cylinders $\mc{M}_1^{\times n}$. The twist field $\Tw_n(x,\tau)$ imposes the boundary conditions $\vphi_k(x',\tau^-)=\vphi_{k+1}(x',\tau^+)$ $\forall_{x'\leq x,\, k}$ with $\vphi_{k+n}\equiv\vphi_{k}$ and, similarly, for $\tilde{\Tw}_n(x,\tau)$ but in reverse order. See Fig.~\ref{fig:Cylinders}b.

The long-range physics of \emph{critical} (and hence gapless) 1d quantum systems are usually captured by a conformal field theory (CFT) \cite{Belavin1984-241,Francesco1997}, characterized by a central charge $c$ \Footnote{Note that for local 1+1d field theories, conformal invariance follows from invariance under Poincar\'{e} and scale transformations \cite{Polchinski1988-383}. Hence, the long-range physics of critical 1d quantum systems with short-range interactions can usually be described by 1+1d CFT.}. The local twist fields then turn out to be so-called primary fields of the CFT with scaling dimension $\Delta_n=\frac{c}{12}(n-1/n)$ \cite{Knizhnik1987-112,Cardy2008-130}. The invariance under conformal transformations is very restrictive and determines two-point functions up to a nonuniversal factor such that $\bra\tilde{\Tw}_n(z)\,\Tw_n(z')\ket_{\CC^{\times n}}=C_n|z-z'|^{-2\Delta_n}$ for $n$ replicas of the complex plane $\CC^{\times n}$ and coordinates $z=x+i\tau$. As detailed in Appx.~\ref{sec:CFT}, one can use this and a conformal mapping from the complex plane to cylinders to evaluate Eq.~\eqref{eq:TwDef} for $L\to\infty$ and large subsystem sizes $\ell\gg\beta$. One finds the power law decay
\begin{equation}\label{eq:Tw}
	\Tr\Dm_A^n=\big\bra\tilde{\Tw}_n(\ell,{\beta}/{2})\,\Tw_n(\ell,0)\big\ket_{\mc{M}_1^{\times n}}
	\to C_n|\beta/\pi|^{-2\Delta_n}
\end{equation}
such that, after analytic continuation from integer $n$ to real $\alpha$,
\begin{subequations}\label{eq:S}
\begin{equation}\label{eq:S_CFT}
	S_\alpha(\ell)\to\frac{c}{6}(1+1/\alpha)\log(\beta/\pi)+C_\alpha'.
\end{equation}
The constants $C_\alpha'=\frac{1}{1-\alpha}\log C_\alpha$ are nonuniversal and depend on the necessary ultraviolet cut-off, which is always present in the considered systems due to the finite lattice spacing. They coincide with constants occurring in ground-state entanglement computations \cite{Calabrese2004,Zhou2005-12}. At zero temperature, the TDS converges to the tensor product of the ground state with itself, $\lim_{\beta\to\infty}|\dmp_\beta\ket=|\mathrm{gs}\ket\otimes|\mathrm{gs}\ket$. This implies that, for finite-size systems, the logarithmic scaling of $S_\alpha(\ell)$ in $\beta$ stops at $\beta^*\propto \ell$ and $S_\alpha(\ell)$ converges to twice the ground-state value, i.e., to $\frac{c}{6}(1+1/\alpha)\log(2\ell/\pi)$. This is in correspondence with the finite-size energy gap $\Delta E\propto 1/L$.
In the context of operator entanglement entropy \cite{Zanardi2001-63,Wang2003-67,Prosen2007-76}, the counterpart of Eq.~\eqref{eq:S_CFT} has been conjectured in Ref.~\cite{Znidaric2008-78} for $\alpha=1$, based on numerics for spin chains.

In \emph{gapped} systems, the energy gap $\Delta E>0$ causes an exponential decay of imaginary-time correlations. Hence,
$\bra\tilde{\Tw}_n(\ell,{\beta}/{2})\,\Tw_n(\ell,0)\ket
\to \bra\tilde{\Tw}_n(\ell,{\beta}/{2})\ket\bra\Tw_n(\ell,0)\ket$ if the correlation length $\xi\propto\Delta E^{-1}$ is much smaller than $\beta/2$. Due to translation invariance in the imaginary-time direction, the correlator, and hence $S_\alpha(\ell)$, will be independent of $\beta$ and, for $\ell,L-\ell\gg\xi$, it is also independent of the subsystem size $\ell$. These considerations can be made more explicit for mass-perturbed CFT (massive relativistic quantum field theory). As shown in Ref.~\cite{Calabrese2004}, one finds $\bra\Tw_n(\ell,0)\ket=C_n\xi^{-\Delta_n}$. Hence, for gapped systems, the R\'{e}nyi entanglement entropies of the TDS saturate as functions of $\beta$,
\begin{equation}\label{eq:S_mass}
	S_\alpha(\ell)\to\frac{c}{6}(1+1/\alpha)\log(\xi)+2C_\alpha',
\end{equation}
\end{subequations}
and grow logarithmically with the correlation length $\xi$.

\begin{figure}[t]
\includegraphics[width=\columnwidth]{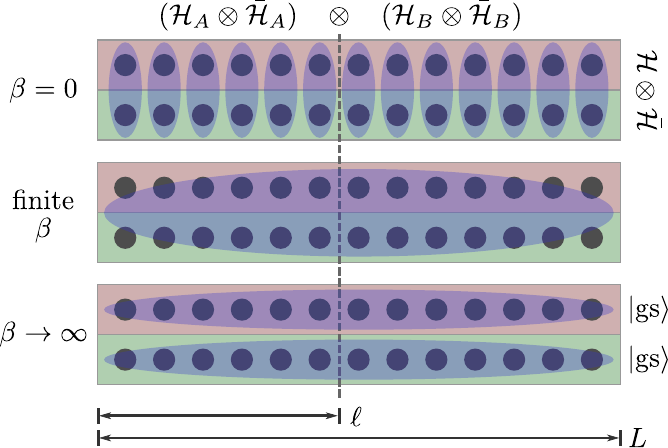}
\caption{\label{fig:TDSentang}\textbf{Structure of TDS entanglement.} The infinite-temperature TDS $|\dmp_0\ket\in\H\otimes\Hb$ has no spatial entanglement. But sites of the original system $\H$ are maximally entangled with corresponding sites of the auxiliary system $\Hb$. During the imaginary-time evolution, spatial entanglement builds up. At zero temperature, the TDS converges to $\lim_{\beta\to\infty}|\dmp_\beta\ket=|\mathrm{gs}\ket\otimes|\mathrm{gs}\ket$. The two copies of the system are then unentangled and, with respect to spatial bipartitions, the TDS has twice the entanglement of the ground state.}
\end{figure}

\section{Complexity of MPS simulations}\label{sec:MPScomplexity}
Following arguments of Verstraete and Cirac for pure states \cite{Verstraete2005-5}, $S_\alpha(\ell)$ can be used to bound the bond dimensions $D$, necessary to approximate the TDS $|\dmp_\beta\ket$ in MPS form \eqref{eq:TDS-MPS} to a certain accuracy. Given a \emph{Schmidt decomposition} $|\dmp_\beta\ket=\sum_k\sqrt{\omega_k(\ell)}\,|k\ket_A\otimes|k\ket_B$ with the $\Dm_A$ eigenvalues $\omega_1\geq \omega_2\geq\dotsc\geq 0$ and orthonormal bases $\{|k\ket_A\}$ and $\{|k\ket_B\}$ for $\H_A\otimes\Hb_A$ and $\H_B\otimes\Hb_B$, one defines the \emph{truncation error} $\epsilon_\ell(D):=\sum_{k>D}\omega_k(\ell)$. As detailed in Appx.~\ref{sec:MPSapprox}, there always exists an MPS approximation with
\begin{equation}\label{eq:MPSboundEps}
	\big(\||\dmp_\beta\ket-|\dmp_\beta^D\ket\|_2\big)^2
	\leq 2\sum_{\ell=1}^{L-1}\epsilon_\ell(D)
\end{equation}
and the truncation error can be bounded by $\log\epsilon_\ell(D)\leq\frac{1-\alpha}{\alpha}[S_\alpha(\ell)-\log(D-1)]$ for $0<\alpha<1$ \cite{Verstraete2005-5}. In comparing systems of variable size $L$, one can consider different requirements for the dependence of $\epsilon_\ell(D)$ on $L$. In the following, let us choose $\epsilon=\epsilon_\ell(D)$ independent of $L$. Solving for bond dimension $D$, we obtain the upper bound
\begin{equation}\label{eq:MPSboundD}
	\log (D-1)\leq S_\alpha(\ell)+\frac{\alpha}{1-\alpha}\log\frac{1}{\epsilon}.
\end{equation}

For fixed $\epsilon$ and $\alpha$, this establishes a system-size independent bound on bond dimensions that is temperature-independent for gapped systems and scales polynomially in the inverse temperature $\beta$ for critical systems such that
\begin{equation}\label{eq:Dscaling}
	D=\mc{O}\left(L^0\beta^0\xi^{\frac{c}{6}\left(1+\frac{1}{\alpha}\right)}\right)\, \text{and}\ 
	D=\mc{O}\left(L^0\beta^{\frac{c}{6}\left(1+\frac{1}{\alpha}\right)}\right)
\end{equation}
according to Eqs.~\eqref{eq:S_mass} and \eqref{eq:S_CFT}, respectively \Footnote{Due to the divergence of the second term in Eq.~\eqref{eq:MPSboundD} for $\alpha\nearrow 1$, knowledge about the scaling behavior of the von Neumann entanglement entropy $\lim_{\alpha\to 1}S_{\alpha}$ is in general \emph{not} sufficient to establish MPS approximability.}. \emph{Thus, 1d quantum systems at finite temperatures can be simulated efficiently on classical computers.} Specifically, the MPS approximation \eqref{eq:TDS-MPS} of the TDS can be computed using an imaginary-time evolution starting at $\beta=0$ with computations costs scaling as $\mc{O}(\beta LD^3)$ \cite{Verstraete2004-6,Schollwoeck2011-326,Barthel2016-94}.
\begin{figure*}[t]
\includegraphics[width=0.82\textwidth]{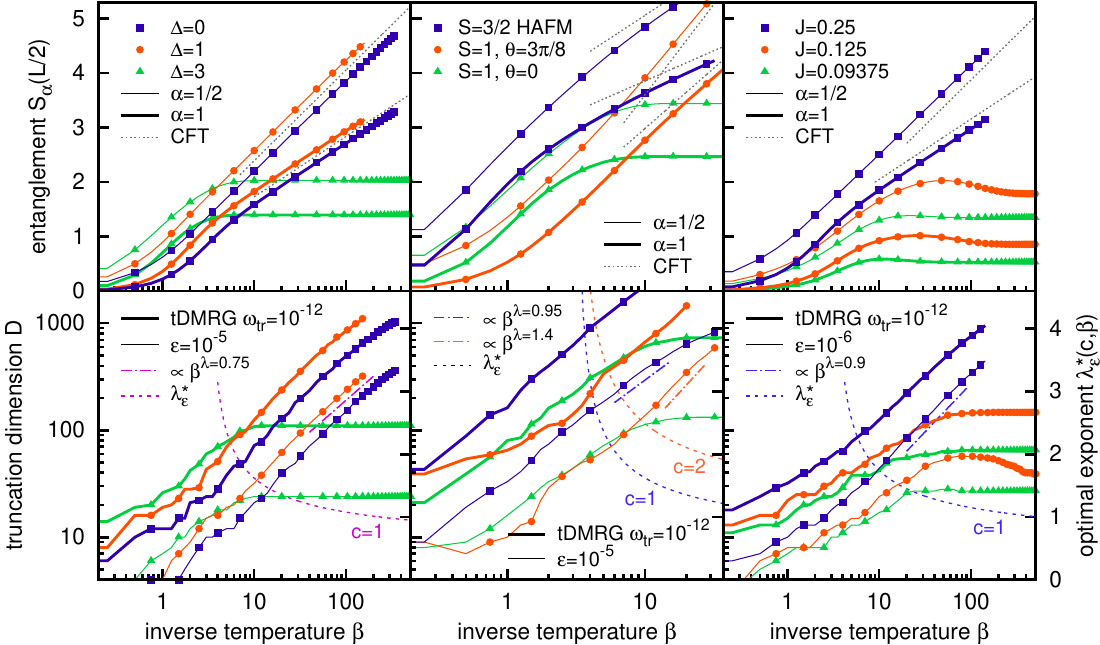}
\caption{\label{fig:Numerics}\textbf{Numerical investigation for spin chains and 1D bosons.} R\'{e}nyi entanglement entropies $S_{\alpha}$ in thermofield double states $|\dmp_\beta\ket$, and truncation dimensions $D$ of corresponding MPS representations. Left: Spin-$1/2$ XXZ chain. Middle: Isotropic spin-$3/2$ Heisenberg antiferromagnet and spin-$1$ bilinear-biquadratic chain. Right: Bose-Hubbard model with onsite repulsion $U=1$, chemical potential $\mu=1/2$, and a maximum of $5$ bosons per site. System sizes $L$ are chosen such that finite-size effects are negligible (either $L=384$ or $L=192$). For critical systems, these quasi-exact results for $S_\alpha(L/2)$ confirm the corresponding CFT prediction \eqref{eq:S_CFT}. The bounds \eqref{eq:MPSboundD} on $D$ are relatively loose, but predicted scaling exponents $\lambda^*=\frac{c}{6}(1+{1}/{\alpha^*})>\frac{c}{3}$ in Eq.~\eqref{eq:Dscaling} with the optimal $\alpha^*$ [Eq.~\eqref{eq:alphaOpt}] compare favorably to numerics.}
\end{figure*}

For $S_\alpha=\frac{c}{6}(1+1/\alpha)\log y$, with $y$ being the correlation length or $\beta$ according to Eqs.~\eqref{eq:S}, we can minimize the bound \eqref{eq:MPSboundD} on the required bond dimension $D$ with respect to $\alpha$. One finds
\begin{equation}\label{eq:alphaOpt}
	\alpha^*=\bigg(1-\sqrt{\frac{6}{c}\log\frac{1}{\epsilon}\big/\log y}\bigg)\Big/
	\bigg(1-\frac{6}{c}\log\frac{1}{\epsilon}\big/\log y\bigg).
\end{equation}
The smallest scaling exponents in Eq.~\eqref{eq:Dscaling} hence occur for large $y$, big central charge $c$, and large truncation error $\epsilon$. For currently typical MPS simulation parameters with $\epsilon\sim 10^{-5}\dots 10^{-14}$, $y~\sim 10\dots 10^3$, and $c\sim 1/2\dots 2$, one has $\alpha^*\sim 0.07\dots 0.3$. Because the bound \eqref{eq:MPSboundD} for $D$ turns out to be relatively loose and in order to test the field-theoretical results \eqref{eq:S} for $S_\alpha$, numerical analyses for spin chain models and the Bose-Hubbard model are provided below.

\section{Relevant norms}\label{sec:Norms}
An alternative to working with MPS purifications $|\dmp\ket$ would be to work with matrix product operator (MPO) representations of $\dm$ \cite{Verstraete2004-6,Zwolak2004-93} which is a reason for studying operator entanglement entropies \cite{Prosen2007-76,Znidaric2008-78}.
Although MPOs are still the best option in some scenarios where suitable purifications are not available, there is an important drawback which is not fully appreciated in the literature. The relevant distance measure for density operators is the \emph{trace distance} $\|\dm-\dm'\|_1\equiv\Tr|\dm-\dm'|=\max_{\hX:\,\|\hX\|=1} \Tr[\hX(\dm-\dm')]$. It provides the maximum difference in expectations values and the success rate for discriminating $\dm$ and $\dm'$ in measurements.

The scaling \eqref{eq:S} guarantees that efficient MPS approximations $|\dmp^D_\beta\ket$ [Eq.~\eqref{eq:TDS-MPS}] of $|\dmp_\beta\ket$ and, analogously, MPO approximations $\dm^D_\beta$ of $\dm_\beta$ are possible in the sense that, with bond dimensions being scaled according to Eq.~\eqref{eq:Dscaling}, the Schatten 2-norm distances are small. However, generically, one only has the bound $\|\hY\|_1\leq\sqrt{\dim\H}\,\|\hY\|_2$ for the trace norm from H\"{o}lder's inequality \cite{Bhatia1997}. Hence, a small distance $\|\dm^D_\beta-\dm_\beta\|_2\leq\varepsilon$ does in general \emph{not} imply a good approximation of $\dm_\beta$ in MPO form unless $\varepsilon$ is chosen to decrease exponentially in the system size $L$. In contrast, one can easily show \Footnote{$\|\dm-\dm'\|_1\leq \||\dmp\ket\bra\dmp|-|\dmp'\ket\bra\dmp'|\|_1=2\sqrt{1-|\bra\dm|\dm'\ket|^2}\leq 2^{3/2}\sqrt{1-|\bra\dm|\dm'\ket|}\leq 2\||\dmp\ket-|\dmp'\ket\|_2$.} that a bound $\||\dmp^D_\beta\ket-|\dmp_\beta\ket\|_2\leq\varepsilon$ for purifications \emph{does} imply $\|\dmt^D_\beta-\dm_\beta\|_1\leq 2\varepsilon$ for the corresponding density operator $\dmt^D_\beta=\Tr_{\Hb}|\dmp^D_\beta\ket\bra\dmp^D_\beta|$.

\section{Numerical investigation}\label{sec:Numerics}
In the following, the field-theoretical formulae \eqref{eq:S} for the R\'{e}nyi entanglement entropies in the TDS $|\dmp_\beta\ket$ and the resulting scaling \eqref{eq:Dscaling} of truncation dimensions $D$ are tested for a few important integrable and non-integrable lattice models, describing, e.g., quantum magnets \cite{Schollwoeck2004magnetism} and ultracold bosonic atoms in optical lattices \cite{Jaksch1998-81,Bloch2007}. Figure~\ref{fig:Numerics} shows results of MPS simulations. The spin-$1/2$ XXZ chain
\begin{equation*}
\hH=\sum_{i=1}^{L-1}(\hS_i^x\hS_{i+1}^x+\hS_i^y\hS_{i+1}^y+\Delta\hS_i^z\hS_{i+1}^z)\end{equation*}
is integrable \cite{Bethe1931,Cloizeaux1966-7,Takahashi1999}. It is critical in the XY phase $-1\leq\Delta\leq 1$ with long-range physics described by Luttinger liquid field theory (the sine-Gordon model) \cite{Luttinger1963-4,Mattis1965-6,Haldane1981-14,Giamarchi2004}
and, hence, a CFT with central charge $c=1$. In the N\'{e}el phase, the gap is $\Delta E=0.613(4)$ at $\Delta=3$.
The non-integrable spin-$3/2$ isotropic Heisenberg antiferromagnet
\begin{equation*}\textstyle
\hH=\sum_{i=1}^{L-1}\hvS_i\cdot\hvS_{i+1}\end{equation*}
is also critical in accordance with Haldane's conjecture \cite{Haldane1983-93,Haldane1983-50} and the Lieb-Schultz-Mattis theorem \cite{Lieb1961,Affleck1986-12,Hastings2004-69}.
The long-range physics of all isotropic odd-integer Heisenberg antiferromagnets can be described by the level-1 $SU(2)$ Wess-Zumino-Witten
model and, hence, a CFT with central charge $c=1$ \cite{Witten1984-92,Affleck1987-36,Fradkin2013}. However, due to the larger number of degrees of freedom, TDS for the spin-$3/2$ chain feature higher entanglement than for spins-$1/2$.
The bilinear-biquadratic spin-$1$ chain
\begin{equation*}
\hH=\sum_{i=1}^{L-1}\big[\cos\theta\,\hvS_i\cdot\hvS_{i+1}+\sin\theta\,(\hvS_i\cdot\hvS_{i+1})^2\big]\end{equation*}
is non-integrable except for special $\theta$ values. In the Haldane phase \cite{Haldane1983-93} at $\theta=0$, it has a gap of $\Delta E=0.410(5)$. It is in a gapless phase for $\pi/4\leq\theta\leq \pi/2$ \cite{Fath1991-44,Fath1993-47,Itoi1997-55,Laeuchli2006-74} with long-range physics governed by the level-1 $SU(3)$ Wess-Zumino-Witten model with marginally irrelevant perturbations \cite{Itoi1997-55}. The corresponding CFT with $c=2$ implies a fast growth of entanglement when temperature is lowered.
Lastly, we consider the Bose-Hubbard model
\begin{equation*}\textstyle
\hH = \sum_{i} \big[-J (\hat{b}^{\dagger}_{i} \hat{b}_{i+1} + \mathrm{h.c.} ) + U \hat{n}_{i}(\hat{n}_{i}-1)/2 - \mu \hat{n}_{i}\big]\end{equation*}
which is also non-integrable. Increasing the hopping $J$ at fixed chemical potential $\mu/U=1/2$, the system undergoes a phase transition from the Mott insulator with one boson per site to a superfluid at $J/U\gtrsim 0.125$ \cite{Fisher1989-40,Kuehner1998-58,Kuehner2000-61,Greiner2002-415}. The superfluid is a Luttinger liquid \cite{Haldane1981-47,Giamarchi2004} and has hence central charge $c=1$. Gaps at $J/U=0.09375, 0.125$ in the Mott phase are $\Delta E/U=0.138(8)$ and $0.027(9)$, respectively.

Figure~\ref{fig:Numerics} shows the results of time-dependent DMRG simulations of the different systems with a 4th order Trotter-Suzuki decomposition \cite{Trotter1959,Suzuki1976}. Bond dimensions $D$ in the MPS $|\dmp_\beta^D\ket$ are adjusted dynamically as in Refs.~\cite{Barthel2013-15,Binder2015-92} by truncating all $\Dm_A$ eigenvalues $\omega_k<\omega_\text{tr}=10^{-12}$ such that R\'{e}nyi entanglement entropies $S_{\alpha=1,1/2}$ in the upper panels of Fig.~\ref{fig:Numerics} are quasi-exact. For the gapped systems, $S_\alpha$ indeed converges at $\beta\sim 1/\Delta E$ in accordance with Eq.~\eqref{eq:S_mass}. For the critical systems, the CFT prediction \eqref{eq:S_CFT} is confirmed for all considered models. (Additive constants for $S_\alpha$ are chosen arbitrarily in these plots.) The predicted scaling of MPS bond dimensions \eqref{eq:Dscaling} is probed in the lower panels. Bond dimensions in the very precise MPS of the actual simulations are shown with thicker lines. From these MPS, we obtain truncation dimensions $D_\epsilon$ (thin lines), corresponding to truncation errors $\epsilon=10^{-5}$ and $10^{-6}$ for the spin systems and bosonic systems, respectively. Scaling exponents $\lambda$ are determined for certain temperature ranges. While the upper bounds \eqref{eq:MPSboundD} are relatively loose, exponents $\lambda$ compare favorably to the predicted values $\lambda^*=\frac{c}{6}(1+{1}/{\alpha^*})>\frac{c}{3}$ with $\alpha^*$ from Eq.~\eqref{eq:alphaOpt}. For example, we find $D_\epsilon\sim\beta^{\lambda=0.75}$ for the critical XXZ model with $\Delta=0,1$ around $\beta=80$  and the bound \eqref{eq:MPSboundD} gives $D_\epsilon=\mc{O}(\beta^{\lambda^*\approx 1})$.

\section{Discussion}
A field-theoretical analysis of the scaling of entanglement entropies in purifications for equilibrium states allowed us to bound the bond dimensions and, hence, the computation costs for corresponding MPS approximations. The analytical results are confirmed by quasi-exact numerical data for several integrable and non-integrable systems. This shows that typical 1d quantum-many body systems \emph{at any temperature} can be simulated efficiently on \emph{classical} computers with costs that are system-size independent and increase, in the worst case (critical systems), only polynomially with the inverse temperature. Consequently, quantum simulations \cite{Feynman1982-21,Lloyd1996-273} are not strictly necessary for this purpose. Nevertheless, it is of course of interest to find quantum protocols with better scaling properties.

I thank H.\ Baranger, M.\ Binder, Y.\ Ikhlef, J.\ Lu, G.\ Schrank, and A.\ Weichselbaum for helpful feedback.

\appendix

\section{Twist-field correlators at criticality}\label{sec:CFT}
To determine R\'{e}nyi entanglement entropies for the TDS \eqref{eq:TDS} within conformal field theory (CFT), we used that, in the limit of infinite system size $L\to\infty$ and subsystem size $\ell\gg\beta$, the two-point correlator of branch-point twist fields obeys $\bra\tilde{\Tw}_n(\ell,{\beta}/{2})\,\Tw_n(\ell,0)\ket_{\mc{M}_1^{\times n}}
\to C_n|\beta/\pi|^{-2\Delta_n}$ [Eq.~\eqref{eq:Tw}],
where $\Delta_n$ is the scaling dimension of the twist fields. This can be sown as follows. On ($n$ copies of) the complex plane, the correlator is simply 
\begin{equation}
 	\big\bra\tilde{\Tw}_n(z_1)\,\Tw_n(z_2)\big\ket_{\CC^{\times n}}=C_n|z_1-z_2|^{-2\Delta_n},
\end{equation}
where $z=x+i\tau$ specifies a point in real space and imaginary time \cite{Belavin1984-241,Francesco1997}.

The conformal transformation $w(z):=\frac{\beta}{2\pi}\log z$ maps the complex plane $\CC$ onto an infinitely long cylinder $\tilde{\mc{M}}_1=w(\CC)$ of circumference $\beta$. Applying the general formula for the transformation of primary fields \cite{Belavin1984-241,Francesco1997}, we obtain with $w_i:=w(z_i)$
\begin{align}
	\big\bra\tilde{\Tw}_n(w_1)\,&\Tw_n(w_2)\big\ket_{\tilde{\mc{M}}_1^{\times n}}\nonumber\\
	&=|w'(z_1)w'(z_2)|^{-\Delta_n} \big\bra\tilde{\Tw}_n(z_1)\,\Tw_n(z_2)\big\ket_{\CC^{\times n}}\nonumber\\
	&=C_n\left|\frac{\beta^2}{(2\pi)^2 z_1z_2}\right|^{-\Delta_n}|z_1-z_2|^{-2\Delta_n}\nonumber\\
	&=C_n\left(\frac{\beta}{2\pi}\right)^{-2\Delta_n} \left|\frac{(z_1-z_2)^2}{z_1z_2}\right|^{-\Delta_n}. \label{eq:TwCorrelator}
\end{align}
Equation~\eqref{eq:Tw} then follows with $w_1=\ell$ and $w_2=\ell+i\beta/2$ which gives $|(z_1-z_2)^2/(z_1z_2)|=4$.

Similarly, we can consider $w_1=x_1\in\mathbb{R}$ and $w_2=x_2\in\mathbb{R}$ which gives $|(z_1-z_2)^2/(z_1z_2)|=4\sinh^2(\pi\,|x_1-x_2|/\beta)$. This then implies that, in the limit $L\to\infty$ and $x_1,x_2\gg\beta$, the finite temperature $\beta^{-1}$ induces a finite correlation length $\xi_\beta$ and spatial correlation functions decay as
\begin{align}
 	\big\bra\phi_1(x_1)\,\phi_2(x_2)\big\ket_{\tilde{\mc{M}}_1}
 	&=\left[\frac{\beta}{\pi}\sinh(\pi\,|x_1-x_2|/\beta)\right]^{-\Delta}\nonumber\\
 	&\sim \left(\frac{\beta}{2\pi}\right)^{-\Delta} e^{-|x_1-x_2|/\xi_\beta}.
\end{align}
Here, $\phi_1$ and $\phi_2$ are primary fields with scaling dimension $\Delta$, and $\xi_\beta=\beta/(\pi\Delta)$ \cite{Belavin1984-241,Francesco1997}. Consequently, our results also hold for finite-size systems, as long as $L\gg \xi_\beta$.

\section{Bounding MPS approximation errors}\label{sec:MPSapprox}
Let $|\psi\ket=\sum_{\vs}\psi^\vs|\vs\ket$ be a normalized many-body state for a system of size $L$ with an orthonormal basis $\{|\vs\ket\equiv|\sigma_1,\dotsc,\sigma_L\ket\}$. For a bipartition of the system into the block $A$, comprising sites 1 to $i$, and block $B$, comprising sites $i+1$ to $L$, let $\omega_i$ be a diagonal matrix that contains the descendingly ordered eigenvalues $\omega_{i,1}\geq\omega_{i,2}\geq\dotsc\geq 0$ of the reduced density matrix $\Dm_A:=\Tr_{\H_B}|\psi\ket\bra\psi|$.

As shown by Verstraete and Cirac in Ref.~\cite{Verstraete2005-5}, for any state $|\psi\ket$, there exists an MPS approximation
\begin{equation}\label{eq:MPS}
	|\psi^D\ket=\sum_{\vs} \tilde{A}_1^{\sigma_1} \tilde{A}_2^{\sigma_2}\dotsb \tilde{A}_L^{\sigma_L}\,|\vs\ket
\end{equation}
with bond dimension $D$ such that truncation errors $\epsilon_i(D):=\sum_{k>D}\omega_{i,k}$ provide a bound on the two-norm distance $\||\psi\ket-|\psi^D\ket\|_2$. Furthermore, the truncation errors can be bounded using R\'{e}nyi entanglement entropies. Let us recapitulate these arguments in the following.

First, note that $|\psi\ket$ can always be represented exactly in MPS form, such that
\begin{equation}\label{eq:MPSexact}
	|\psi\ket=\sum_{\vs}\psi^\vs|\vs\ket
	=\sum_{\vs} A_1^{\sigma_1} A_2^{\sigma_2}\dotsb A_L^{\sigma_L}\,|\vs\ket
\end{equation}
with bond dimensions growing exponentially in $L$. This can be achieved by a sequence of singular value decompositions (SVD) \cite{Golub1996} that decompose the tensor $\psi^\vs$ into the product $\prod_i A_i^{\sigma_i}$. The SVDs correspond to Schmidt decompositions \cite{Nielsen2000} of $|\psi\ket$, e.g., for a cut at bond $(i,i+1)$ such that the resulting singular value spectrum is given by $\sqrt{\omega_i}$. Considering $\psi^\vs$ as a matrix with row multi-index $(\sigma_1,\dotsc,\sigma_{L-1})$ and column index $\sigma_L$, the first SVD yields $\psi^\vs\stackrel{\SVD}{=:}\psi_{L-1}^{\sigma_1,\dotsc,\sigma_{L-1}}\sqrt{\omega_{L-1}}A_L^{\sigma_L}$. We then continue with
\begin{equation}\label{eq:SchmidtRecurr}
	\psi_i^{\sigma_1,\dotsc,\sigma_i}\sqrt{\omega_{i}}
	\stackrel{\SVD}{=:}\psi_{i-1}^{\sigma_1,\dotsc,\sigma_{i-1}}\sqrt{\omega_{i-1}}A_i^{\sigma_i} 
\end{equation}
for $i=L-1,\dotsc,1$.
The resulting matrices $A_i^{\sigma}$ obey the (right) orthonormality constraints
\begin{equation}\label{eq:rightON}
	\sum_\sigma A_i^\sigma(A_i^\sigma)^\dag=\id
	\quad\text{and}\quad
	\sum_\sigma (A_i^\sigma)^\dag\omega_{i-1}A_i^\sigma=\omega_{i}
\end{equation}
which is in accordance with $A_i$ playing the role of a linear isometry (projection onto a subspace) in the traditional view of the DMRG method \cite{White1992-11,White1993-10,Rommer1997,Schollwoeck2005}. The properties \eqref{eq:rightON} follow from the recurrence relation \eqref{eq:SchmidtRecurr}. In particular, the second part follows from,
\begin{align*}
	\omega_i&\,\,=\sum_{\vs',\sigma_i}\sqrt{\omega_i}(\psi^{\vs',\sigma_i}_i)^\dag \psi^{\vs',\sigma_i}_i\sqrt{\omega_i}\\
	&\stackrel{\text{\eqref{eq:SchmidtRecurr}}}{=} \sum_{\vs',\sigma_i} (A_i^{\sigma_i})^\dag \sqrt{\omega_{i-1}}(\psi_{i-1}^{\vs'})^\dag\psi_{i-1}^{\vs'}\sqrt{\omega_{i-1}}A_i^{\sigma_i}\\
	&\,\,=\sum_{\sigma_i} (A_i^{\sigma_i})^\dag\omega_{i-1}A_i^{\sigma_i},
\end{align*}
where $\vs'=(\sigma_1,\dotsc,\sigma_{i-1})$ and property $\sum_{\vs} (\psi^{\vs}_i)^\dag \psi^{\vs}_i=\id$ has been used twice.

Now, let us employ the obtained exact representation \eqref{eq:MPSexact} to define an MPS approximation $|\psi^D\ket$ with bond dimension $D$. In particular, choose $\tilde{A}^{\sigma}_i$ in Eq.~\eqref{eq:MPS} as the upper left $D\times D$ block of $A^{\sigma}_i$, i.e., $\tilde{A}^{\sigma}_i:=P A^{\sigma}_i P$ for $1<i<L$, where projector $P$ has matrix elements $[P]_{a,b}=\sum_{k=1}^D\delta_{a,k}\delta_{b,k}$. Correspondingly, $\tilde{A}^{\sigma}_1:=A^{\sigma}_1 P$ and $\tilde{A}^{\sigma}_L:=P A^{\sigma}_L$. We will see that this implies the bound 
\begin{equation}\label{eq:MPSerrorBound}
	\||\psi\ket-|\psi^D\ket\|_2^2\leq 2\sum_{i=1}^{L-1}\epsilon_i(D)
\end{equation}
for the two-norm distance \cite{Verstraete2005-5}. According to Eq.~\eqref{eq:rightON}, $\E_i(X):=\sum_\sigma (A_i^\sigma)^\dag X A_i^\sigma$ is a quantum channel \cite{Wilde2017} with Kraus operators $\{(A_i^\sigma)^\dag\}$ and the property $\E_i(\omega_{i-1})=\omega_i$. This allows us to write the overlap $\bra\psi|\psi^D\ket$ as
\begin{align*}
	\bra\psi|\psi^D\ket
	&=\sum_\vs (A_L^{\sigma_L})^\dag\dotsb (A_1^{\sigma_1})^\dag \tilde{A}_1^{\sigma_1} \dotsb \tilde{A}_L^{\sigma_L}\\
	&=\E_L(\dots\E_3(\E_2(\omega_1 P)P)\dots P).
\end{align*}
As discussed in Appx.~\ref{sec:traceNorm}, the trace norm $\|X\|_1:=\Tr\sqrt{XX^\dag}$ is non-increasing under application of quantum channels, $\|\E_i(X)\|_1\leq \|X\|_1$, and also non-increasing with respect to insertion of projectors, $\|XP\|_1\leq\|X\|_1$.
This can be used to bound $|1-\bra\psi|\psi^D\ket|$. In particular, with $\omega_L=1$, $Y_1:=\omega_1$, and $Y_i:=\E_i(Y_{i-1}P)$ such that $Y_L=\bra\psi|\psi^D\ket$,
\begin{align*}
	|1-\bra&\psi|\psi^D\ket|
	=\|\omega_L-Y_L\|_1
	=\|\E_L(\omega_{L-1}-Y_{L-1}P)\|_1\\
	&\leq \|\omega_{L-1}-Y_{L-1}P\|_1\\
	&\leq \|(\omega_{L-1}-Y_{L-1})P\|_1+\|\omega_{L-1}\,(\id-P)\|_1\\
	&\leq\|\omega_{L-1}-Y_{L-1}\|_1+\epsilon_{L-1}(D)
	=\dotsc=\sum_{i=1}^{L-1}\epsilon_i(D),
\end{align*}
where we have used that $\|\omega_i\,(\id-P)\|_1=\sum_{k>D}\omega_{i,k}$ is simply the truncation error $\epsilon_i(D)$. Similarly, as also $\|PXP\|_1\leq \|X\|_1$, 
\begin{align*}
	\bra\psi&^D|\psi^D\ket=\E_L(P\E_{L-1}(P\dots\E_{2}(P\omega_1 P)\dots P) P)\\
	&\leq \|\E_{L-1}(P\dots\E_{2}(P\omega_1 P)\dots P)\|_1
	\leq\dotsc \leq\|\omega_1\|_1=1.
\end{align*}
With $\big(\||\psi\ket-|\psi^D\ket\|_2\big)^2\leq 2|1-\bra\psi|\psi^D\ket|+\bra\psi^D|\psi^D\ket-1$,
this proves the bound \eqref{eq:MPSerrorBound}.

Finally, one can relate the truncation error $\epsilon(D)$ for a cut at some bond $(i,i+1)$ to the R\'{e}nyi entanglement entropy $S_\alpha(\omega)=\frac{1}{1-\alpha}\log\sum_k(\omega_k)^\alpha$ \cite{Verstraete2005-5}, where  $\omega_1\geq\omega_2\geq\dotsc\geq 0$ ($\sum_k\omega_k=1$) denote the decreasingly ordered eigenvalues of the reduced density matrix $\Dm_A$ of subsystem $A$, which comprises lattice sites $1$ to $i$. $S_\alpha$ is \emph{Schur-concave} \cite{Bhatia1997}. This means that $S_\alpha(\omegat)\leq S_\alpha(\omega)$ if the distribution $\omegat$ \emph{majorizes} $\omega$, i.e., if $\sum_{k=1}^n\omegat_k\geq \sum_{k=1}^n\omega_k$ $\forall_n$. Hence, with a given truncation error $\epsilon(D)=\sum_{k>D}\omega_k$, we can obtain a lower bound for $S_\alpha(\omega)$ by finding a distribution $\omegat$ that majorizes all distributions with the same truncation error $\epsilon=\epsilon(D)$. With a second positive parameter $h=\omegat_D$, let us choose
\begin{align}\label{eq:optimalDistrib}
\begin{split}
	&\omegat_1=1-\epsilon-(D-1)h,\quad
	\omegat_2=\dotsc=\omegat_{D+K}=h,\\
	&\omegat_{D+K+1}=\gamma h,\quad\text{and}\quad
	\omegat_{k}=0\ \ \forall_{k>D+K},
\end{split}
\end{align}
where $K=\lfloor\epsilon/h\rfloor$ and $\gamma=\epsilon/h-K\in[0,1)$. This distribution has truncation error $\sum_{k>D}\omegat_k=\epsilon$ and majorizes all other distributions $\omega'$ with the same truncation error and $\omega'_D=h$ as
\begin{equation*}
	\sum_{k=1}^n\omegat_k=
	\begin{cases}
	 1-\epsilon-(D-n)h & \text{for}\ 1\leq n\leq D+K\\
	 1 & \text{for}\ n>D+K
	\end{cases}
\end{equation*}
and $\sum_{k=1}^n\omega'_k=1-\epsilon-\sum_{k=n+1}^D\omega_k'\leq 1-\epsilon-(D-n)h$ for $1\leq n\leq D$, $\sum_{k=1}^n\omega'_k=1-\epsilon+\sum_{k=D+1}^n\omega_k'\leq 1-\epsilon-(D-n)h$ for $D< n\leq D+K$ and, of course, $\sum_{k=1}^n\omega'_k\leq 1$ for $n>D+K$. In a final step, we need to minimize $S_\alpha(\omegat)$ with respect to $h$ to find the globally optimal distribution for the given $\epsilon$ and $D$. While this is not easily possible, it is simple to minimize the lower bound $\frac{1}{1-\alpha}\log\left[(D-1+\epsilon/h)h^{\alpha}\right]\leq S_\alpha(\omegat)\leq S(\omega)$ with respect to $h$. One obtains $h^*=\frac{1-\alpha}{\alpha}\frac{\epsilon}{D-1}$ and hence \cite{Verstraete2005-5}
\begin{align}\label{eq:SboundEps}
\begin{split}
	S_\alpha(\omega)
	&\geq \frac{1}{1-\alpha}\log\frac{(D-1)^{1-\alpha}\epsilon^\alpha}{\alpha^\alpha(1-\alpha)^{1-\alpha}}\\
	&\geq \frac{1}{1-\alpha}\log\left[(D-1)^{1-\alpha}\epsilon^\alpha\right]
\end{split}
\end{align}
which is valid as long as
\begin{equation}\label{eq:alphaRange}
	\omegat_1\geq\omegat_2=h^*>0 \ \ \Leftrightarrow\ \
	1>\alpha\geq\frac{\epsilon D}{D-1-\epsilon}\approx \epsilon.
\end{equation}
If a R\'{e}nyi entanglement entropy $S_\alpha$ with $\alpha$ in the range \eqref{eq:alphaRange} is known, Eq.~\eqref{eq:SboundEps} can be used to bound $\epsilon$ from above for a given $D$ or to bound $D$ from above for a given desired approximation accuracy [cf.\ Eq.~\eqref{eq:MPSerrorBound}].

In comparison to Ref.~\cite{Verstraete2005-5}, we have $D-1$ instead of $D$ in the bound \eqref{eq:SboundEps}. This minor difference is due to taking into account that $\omegat_{D+K+1}$ can in general not be set to $h$ or $0$.

\section{Trace norm and quantum channels}\label{sec:traceNorm}
Quantum channels $\E:\mc{L}(\H)\to\mc{L}(\H')$ are linear completely positive trace-preserving maps between operator spaces. For every quantum channel, there exists a linear isometry $\hat{V}:\H\to\H'\otimes\H_\aux$ ($\hat{V}^\dag\hat{V}=\hid_\H$) with an auxiliary Hilbert space $\H_\aux$ such that $\E(\hX)=\Tr_\aux \hat{V}\hX\hat{V}^\dag$. This is Stinespring's dilation theorem \cite{Stinespring1955-6}.

The trace norm $\|\hX\|_1\equiv\Tr\sqrt{\hX^\dag \hX}$ (or ``Schatten 1-norm'') can be obtained by maximizing over unitaries in the sense that $\|\hX\|_1=\max_{\hU:\,\hU^\dag\hU=\id}|\Tr\hX\hU|$ \cite{Wilde2017}. It is hence non-increasing under partial traces. As the trace norm gives the sum of the singular values of $\hX$, it is also invariant under isometric transformations. Consequently, the trace norm is non-increasing under application of quantum channels, $\|\E(\hX)\|_1=\|\Tr_\aux \hat{V}\hX\hat{V}^\dag\|_1\leq\|\hat{V}\hX\hat{V}^\dag\|_1=\|\hX\|_1$.

For projection operators $\hat{P}=\hat{P}^\dag=\hat{P}^2$, H\"{o}lder's inequality \cite{Bhatia1997} tells us that $\|\hX\hat{P}\|_1\leq \|\hX\|_1 \|\hat{P}\|_\infty=\|\hX\|_1$ and of course also $\|\hat{P}\hX\hat{P}\|_1\leq\|\hX\|_1$.

The Schatten 2-norm $\|\hX\|_2:=(\Tr\hX^\dag\hX)^{1/2}$ for an operator on $\H$ is typically not very useful for bounding its trace norm $\|\hX\|_1$. Applying H\"{o}lder's inequality, one only gets $\|\hX\|_1\leq \|\hid\|_2\|\hX\|_2=\sqrt{\dim\H}\,\|\hX\|_2$.

\bibliographystyle{prsty.tb.title}

\end{document}